\documentclass[a4paper,12pt]{article}
\usepackage{amsmath,amssymb,bm}
\usepackage[utf8x]{inputenc}
\usepackage{textcomp,marvosym}
\usepackage{xcolor}
\usepackage{appendix}
\usepackage{siunitx} 
\usepackage{booktabs} 

\usepackage{graphicx}

\newcommand{\includetikz}[1]{%
    \includegraphics{#1.pdf}
}

\newlength\figureheight
\newlength\figurewidth
\newcommand{\includetikzfigure}[4]{%
    \setlength\figurewidth{#2\columnwidth}
    \setlength\figureheight{#3\columnwidth}
    \begin{figure}[!ht]
    \centering
    \includetikz{#1}
    \caption{#4}
    \label{fig:#1}
    \end{figure}
}

\newcommand{\citep}{\cite}
\newcommand{\citet}{\cite}

\title{From biophysical to integrate-and-fire~modelling}
\date{}
\author{Tomas Van Pottelbergh \and Guillaume Drion \and Rodolphe Sepulchre}

\begin{document}

\maketitle

\section{Abstract}
\label{sec:abstract}

This paper proposes a methodology to extract a low-dimensional integrate-and-fire model from an arbitrarily detailed single-compartment biophysical model. The method aims at relating the modulation of maximal conductance parameters in the biophysical model to the modulation of parameters in the proposed integrate-and-fire model. The approach is illustrated on two well-documented examples of cellular neuromodulation: the transition between Type I and Type II excitability and the transition between spiking and bursting.

\section{Introduction}
\label{sec:introduction}

Integrate-and-fire models have a long history, dating back to the beginning of the 20th century \citep{Lapicque1907}. Because of their simplicity, they have long served as phenomenological models for action potential generation in neurons. Over the last few decades, the original Leaky Integrate-and-Fire model \citep{Hill1936,Stein1965,Knight1972} has gradually been extended and modified to explain more neurological data and phenomena. Important examples include the replacement of the linear function by a quadratic or exponential nonlinearity \citep{Ermentrout1996,Latham2000,Fourcaud-Trocme2003}, the linearisation of the subthreshold dynamics in biophysical models \citep{Mauro1970,Brunel2014}, and/or adding additional variables modelling the effects of refractoriness and adaptation \citep{Wehmeier1989,Mihalas2008,Jolivet2006,Mensi2012,Pozzorini2013,Izhikevich2003,Brette2005,Drion2012}. Low-dimensional integrate-and-fire models have also been shown to be sufficient to reproduce experimental spike trains with great accuracy \citep{Jolivet2004,Badel2008}.

The simplicity of integrate-and-fire models nevertheless comes with an important limitation: they lack the physiological interpretation of a biophysical model \citep{VanPottelbergh2018}. This is a severe obstacle when studying the role of cellular neuromodulation in a large network \citep{Drion2018}. While intrinsic neuromodulation is studied via changes of maximal conductances in biophysical models, mapping this effect in an integrate-and-fire model with no biophysical connection is challenging.

The objective of the present paper is to provide a quantitative computational bridge between a detailed single-compartment biophysical model and its compact representation by a multi-scale integrate-and-fire model. Our motivation is to allow for a systematic mapping between the physiological parameters of a biophysical model and the abstract parameters of its integrate-and-fire approximation. Our models are not optimised to fit experimental spiking data. Instead, we aim at simplified models amenable to qualitative phase-portrait analysis, and able to capture the qualitative changes of neuronal excitability induced by neuromodulators, which requires nonlinear subthreshold dynamics.

We exploit the architecture of the Multi-Quadratic Integrate-and-Fire (MQIF) model introduced in \citet{Drion2012} and further studied in \citet{VanPottelbergh2018}. The state variables of the model have the interpretation of the membrane potential filtered in different timescales. The model parameters relate to a local balance between positive and negative conductance\footnote{These conductances are not absolute, but differential conductances, i.e. slopes of the I-V curve. The negative conductance is sometimes referred to as ``negative slope conductance'' in the literature, e.g. in the early observation of \citet{Stafstrom1982}. See also \citet{Ceballos2017} for a recent review.} in each timescale. This model has been quite successful at capturing important modulation properties in a robust and qualitative manner.

We introduce a multi-scale integrate-and-fire model in Section~\ref{sec:VC-analysis-reduced-model} and highlight a key feature of the proposed approach: fitting the parameters from voltage-clamp data rather than from current-clamp data. The parameters of this integrate-and-fire model to be identified from the biophysical model can be split into two groups. Section~\ref{sec:VC-analysis-ident-i-ion-1} discusses how to identify the ion current function of the integrate-and-fire model. The other parameters are treated in Section~\ref{sec:VC-analysis-init-opt}, which also discusses local optimisation to improve the result. We then apply this method to model modulation in two biophysical models from the literature in Section~\ref{sec:VC-analysis-PP-analysis}. We end by discussing the limitations of the method and its connection to other methods for the analysis of neural behaviour.

\section{A multi-scale integrate-and-fire model structure}
\label{sec:VC-analysis-reduced-model}

We consider a general multi-scale neuronal model of the form
\begin{align}
    C\dot{V} &= I_{\text{app}} - I_{\text{ion}}(V,V_s,V_{us},\ldots) \label{eq:VC-analysis-reduced-model-1}\\
    \tau_s \dot{V_s} &= V - V_s \label{eq:VC-analysis-reduced-model-2}\\
    \tau_{us} \dot{V}_{us} &= V - V_{us} \label{eq:VC-analysis-reduced-model-3}\\
    & \ldots \nonumber
\end{align}

The voltage equation \eqref{eq:VC-analysis-reduced-model-1} is the classical Kirchhoff relationship of a biophysical model: the current $I_{\text{ion}}$ is the total ionic current intrinsic to the cellular membrane composition. Its voltage-dependence is modelled through the voltage variable $V$ and lagged variables ($V_s$, $V_{us}$, \dots) that model the voltage filtered in distinct timescales (slow, ultraslow, \dots). Those variables differ from the gating variables of traditional conductance based models, but closely relate to the equivalent potentials originally defined in \citet{Kepler1992}. Note that the dynamics of those filters is chosen to be linear. The nonlinearity of the model is entirely concentrated in the scalar function $I_{\text{ion}}$.

While the model \eqref{eq:VC-analysis-reduced-model-1}--\eqref{eq:VC-analysis-reduced-model-3} can exhibit highly nonlinear behaviour in current clamp, this is not the case in voltage clamp. The voltage-clamped relation from $V$ to $I_{\text{app}}$ is shown in Figure~\ref{fig:Wiener-model}. It has the classical structure of a parallel Wiener system. The identification of this nonlinear fading memory model is much simpler than the direct identification of the current clamp dynamics \citep{Burghi2019}. In particular, our methodology identifies the nonlinear function $I_{\text{ion}}$ from the response of the biophysical model to voltage step inputs.

\includetikzfigure{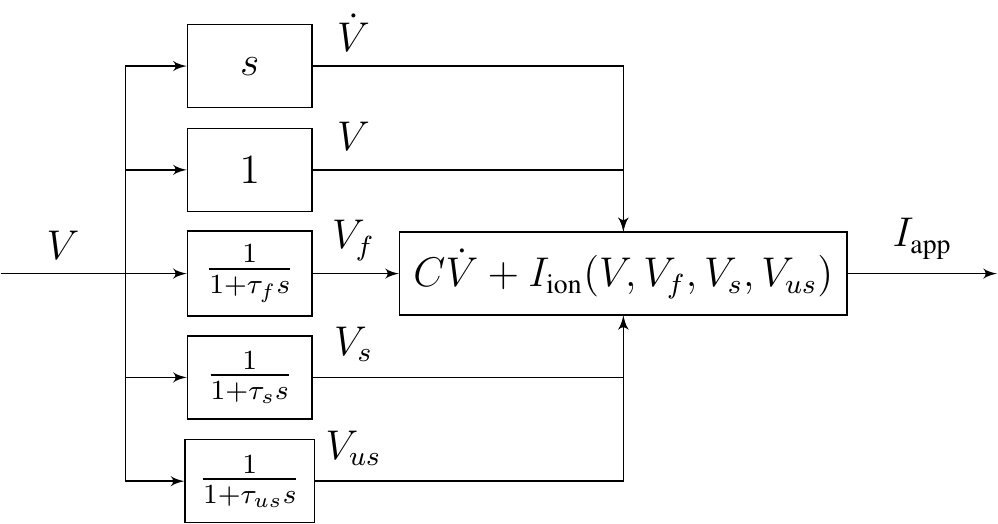}{1}{1}{Block diagram representation of the multi-scale model \eqref{eq:VC-analysis-reduced-model-1}--\eqref{eq:VC-analysis-reduced-model-3} in voltage clamp. Note the additional voltage $V_f$, accounting for the dynamics of the fast ion channels. While this voltage will be merged with $V$ in the final multi-scale model, $\tau_f$ is necessary for the identification of $I_{\text{ion}}$ (Section~\ref{sec:VC-analysis-ident-i-ion-1}). This is a parallel Wiener system with the linear systems on the left feeding into the nonlinearity on the right.}

When a reset is added to the multi-scale model, as in~\eqref{eq:VC-analysis-NLIF-1}--\eqref{eq:VC-analysis-NLIF-3}, the model is converted into an integrate-and-fire model. This simple integrate-and-fire model structure includes many models in the literature: the one-dimensional leaky, quadratic and exponential integrate-and-fire models, but also the multi-dimensional Izhikevich \citep{Izhikevich2003}, AdEx \citep{Brette2005}, and Multi-Quadratic Integrate-and-Fire models \citep{Drion2012,VanPottelbergh2018}.

\begin{align}
  & & \text{if }V &\ge V_\text{max}: \nonumber\\
  C\dot{V} &= I_{\text{app}} - I_{\text{ion}}(V,V_s,V_{us},\ldots) \label{eq:VC-analysis-NLIF-1} & V &\gets V_{r}\\
  \tau_s\dot{V}_s &= V-V_s & V_s &\gets V_{s,r} \label{eq:VC-analysis-NLIF-2}\\
  \tau_{us} \dot{V}_{us} &= V - V_{us} & V_{us} &\gets V_{us} + \Delta V_{us} \label{eq:VC-analysis-NLIF-3}\\
    & \ldots \nonumber
\end{align}

In the remainder of this paper, we concentrate on the integrate-and-fire version of the model. The reader should however be aware that there is a direct correspondence between the model \eqref{eq:VC-analysis-reduced-model-1}--\eqref{eq:VC-analysis-reduced-model-3} and the model \eqref{eq:VC-analysis-NLIF-1}--\eqref{eq:VC-analysis-NLIF-3} provided that the model \eqref{eq:VC-analysis-reduced-model-1}--\eqref{eq:VC-analysis-reduced-model-3} generates spikes. Each spike in the continuous-time model is replaced by a reset mechanism in the model \eqref{eq:VC-analysis-NLIF-1}--\eqref{eq:VC-analysis-NLIF-3}.

The motivation for this substitution is computational: the reset avoids the stiff integration of a spike, which can be a significant computational gain in the numerical integration of a large-scale spiking model. Instead, the numerical integration is stopped at threshold and reinitialised at a novel initial condition specified by a discrete rule. We insist that the reset operation in our integrate-and-fire model has no other role than short-cutting the (otherwise tedious) numerical integration of the spike in the differential equation. This is in contrast with other integrate-and-fire models of the literature that design the reset rule to generate solutions that do not correspond to solutions of the differential equation (see \citet{VanPottelbergh2018} for details). This constraint is key to retain a direct correspondence to the physiology of a biophysical model.

An important property of the integrate-and-fire model \eqref{eq:VC-analysis-NLIF-1}--\eqref{eq:VC-analysis-NLIF-3} is that the nonlinear function $I_{\text{ion}}$ only needs to be identified in the subthreshold voltage range $V < V_\text{max}$.

\section{Identification of the ion current function from a biophysical model}
\label{sec:VC-analysis-ident-i-ion-1}

Our methodology approximates a given biophysical model by an integrate-and-fire model of the type \eqref{eq:VC-analysis-NLIF-1}--\eqref{eq:VC-analysis-NLIF-3}. We make a distinction between the ``structural'' parameters of the model (the time constants and the reset parameters) and the single scalar nonlinear function $I_{\text{ion}}$. In this section, we determine $I_{\text{ion}}$ for a given set of structural parameters. In the next section, we address the determination of the structural parameters and how the design can be optimised by iterating between the identification of the total ionic current and the identification of the structural parameters.

Given a biophysical model and choice of structural parameters, we propose to identify the ionic current $I_{\text{ion}}$ such as to optimise the matching between voltage-clamp experiments on the biophysical model and on the integrate-and-fire approximation. We choose this criterion because it combines physiological relevance and computational tractability. It is physiologically relevant because biophysical models are developed from voltage-clamp experiments in the first place. It is also computationally tractable because a voltage-clamp step response can be calculated in closed form both in a biophysical model and in the integrate-and-fire model. In a biophysical model, each gating variable obeys a differential equation of the form
\begin{equation}
     \tau_{x_i}(V) \dot{x}_i = x_{i,\infty}(V) - x_i,
\end{equation}
which becomes linear for a fixed value of $V$. The solution after a step from $V_0$ to $V_{\text{step}}$ at $t=0$ is described by
\begin{equation}
    x_i(t) = x_{i,\infty}(V_0)+[x_{i,\infty}(V_{\text{step}})-x_{i,\infty}(V_0)]\cdot \left[1-e^{-t/\tau_{x_i}(V_{\text{step}})}\right], \label{eq:VC-analysis-x-sub}
\end{equation}
assuming the membrane potential is at equilibrium at $t=0$.

The response of the total ionic current $I_{\text{ion}}$ in the biophysical model to a voltage-clamp step is then obtained by direct substitution of $V(t)=V_{\text{step}}$ and $x_i(t)$ (given by \eqref{eq:VC-analysis-x-sub}) in the expression for the total ionic current. The same calculation holds in the integrate-and-fire model \eqref{eq:VC-analysis-NLIF-1}--\eqref{eq:VC-analysis-NLIF-3} to obtain an expression of $I_{\text{ion}}(V(t),V_s(t),V_{us}(t),\dots)$.

The details of this simple idea are provided in the next sections.

\subsection{Two-timescale models}
\label{sec:VC-analysis-ident-i-ion-1-2ts}

We start by describing the method for biophysical models that can be described well by a fast and slow timescale, which are well-separated ($\tau_s \gg \tau_f$). We consider a simple voltage clamp step experiment as in Figure~\ref{fig:VC-step}, stepping from the initial voltage $V_0$ to the final voltage $V_{\text{step}}$. Assuming $\tau_f$ is a good approximation of the time constants of the fast gating variables, they will have approximately reached their steady-state value $x_{i,\infty}(V_{\text{step}})$ at $t=3\tau_f$. At the same time, because of timescale separation, the slow gating variables will not have significantly changed from their steady-state value $x_{i,\infty}(V_0)$.

\includetikzfigure{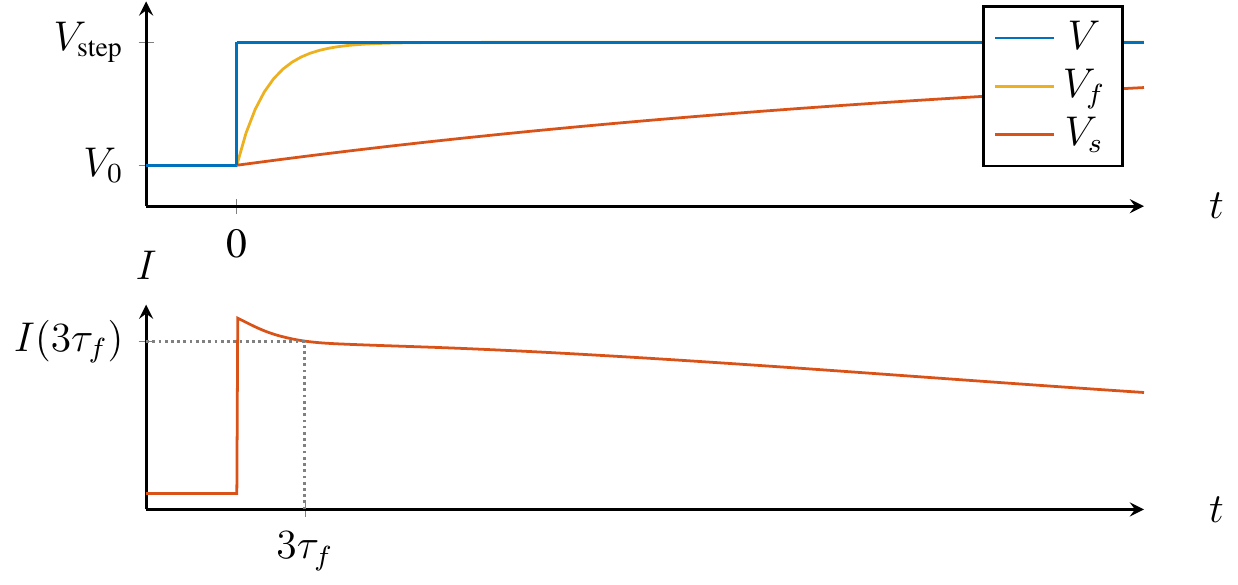}{.8}{.25}{Voltage clamp step experiment on a two-timescale model (the Connor-Stevens model \citep{Connor1971,Connor1977}). The voltage is stepped from $V_0$ to $V_{\text{step}}$. Assuming $V=V_f$, the value of $I$ at $3\tau_f$ will give a good approximation of $I_{\text{ion}}(V_{\text{step}},V_0)$ of the multi-scale model.}

Similarly, in the multi-scale integrate-and-fire model, the fast voltage $V_f$ will have approximately reached its steady-state value $V_{\text{step}}$ at $t=3\tau_f$. Under the assumption of $\tau_s \gg \tau_f$, the slow voltage $V_s$ will still be approximately equal to $V_0$. After merging the variables $V$ and $V_f$ in the multi-scale model, the following observation can be made for the voltage-clamp experiment in Figure~\ref{fig:VC-step}:
\begin{align}
    I(3\tau_f) \approx I_{\text{ion}}(V_{\text{step}},V_0).
\end{align}

The interpretation of this expression is that the value of the ion current function of the multi-scale model evaluated at $V=V_{\text{step}}$ and $V_s=V_0$ can be read of at $t=3\tau_f$ from a voltage-clamp step experiment from $V_0$ to $V_{\text{step}}$. Given that $V_0$ and $V_{\text{step}}$ can be chosen arbitrarily, this gives a simply method to evaluate $I_{\text{ion}}$ on the biophysical model and use its value in the function $I_{\text{ion}}(V,V_s)$ of the multi-scale model.

Instead of doing this matching using actual voltage-clamp experiments, we can use the analytical solution for voltage-clamp steps on biophysical models \eqref{eq:VC-analysis-x-sub}, resulting in the expression
\begin{align}
    \label{eq:VC-analysis-x-sub-2ts}
    x_i = x_{i,\infty}(V_0)+[x_{i,\infty}(V_{\text{step}})-x_{i,\infty}(V_0)]\cdot [1-e^{-3\tau_f/\tau_{x_i}(V_{\text{step}})}]
\end{align}
for each gating variable. Substituting this expression in the ion current equation of the biophysical model then makes $I_{\text{ion}}(V_{\text{step}},V_0)$ of the integrate-and-fire model a function of the equations for the biophysical model and the time constant $\tau_f$. It is clear that in the limit of infinite timescale separation, the formula becomes a simple substitution of the fast and slow gating variables by their steady-state values at $V_{\text{step}}$ and $V_0$ respectively:
\begin{align}
    \lim_{\tau_{x_i} \to 0} x_i &= x_{i,\infty}(V_{\text{step}})\\
    \lim_{\tau_{x_i} \to +\infty} x_i &= x_{i,\infty}(V_0).
\end{align}

\subsection{Three-timescale models}
\label{sec:VC-analysis-ident-i-ion-1-3ts}

We apply the same idea to three-timescale models to find $I_{\text{ion}}(V,V_s,V_{us})$ as a function of the solution of a voltage clamp experiment. The simple procedure for two-timescale models cannot be used anymore, as it would always couple the slow and ultraslow voltage, and therefore only evaluate $I_{\text{ion}}(V_{\text{step}},V_0,V_0)$. This can be resolved by devising a slightly more complex voltage clamp step experiment (see Figure~\ref{fig:VC-step-3ts}): starting at $V_0$, stepping to $V_{\text{step},1}$ and finally to $V_{\text{step},2}$ after $3\tau_s$.

\includetikzfigure{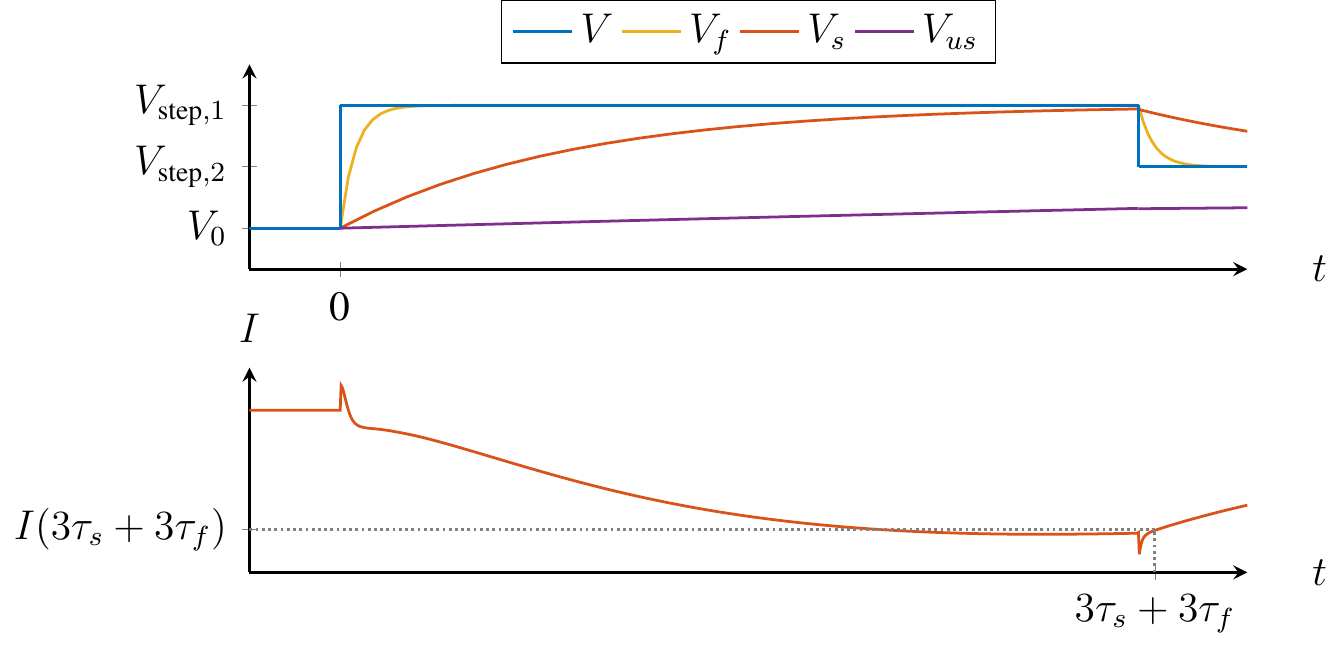}{.8}{.25}{Voltage clamp step experiment on a three-timescale model (the model of \citet{Drion2018}). The voltage is stepped from $V_0$ to $V_{\text{step},1}$ and then to $V_{\text{step},2}$ after $3\tau_s$. Assuming $V=V_f$, the value of $I$ at $3\tau_s+3\tau_f$ will give a good approximation of $I_{\text{ion}}(V_{\text{step},2},V_{\text{step},1},V_0)$ of the multi-scale model.}

Following the same reasoning as before, after $3\tau_s+3\tau_f$ the fast gating variables will be approximately at their steady-state value $x_{i,\infty}(V_{\text{step},2})$. Similarly will the slow gating variables approximately be at the steady-state value $x_{i,\infty}(V_{\text{step},1})$, while the ultraslow gating variables will still be close to $x_{i,\infty}(V_0)$. The value of the current at time $3\tau_s+3\tau_f$ will therefore be a good approximation of $I_{\text{ion}}(V_{\text{step},2},V_{\text{step},1},V_0)$ of the multi-scale model, assuming $V=V_f$.

The necessary substitution for the gating variables in the ion current equation is given by
\begin{align}
    x_i &= \bar{x}_i + [x_{i,\infty}(V_{\text{step},2})-\bar{x}_i]\cdot [1-e^{-3\tau_f/\tau_{x_i}(V_{\text{step},2})}] \label{eq:VC-analysis-x-sub-3ts-1}\\
    \bar{x}_i &= x_{i,\infty}(V_0)+[x_{i,\infty}(V_{\text{step},1})-x_{i,\infty}(V_0)]\cdot [1-e^{-3\tau_s/\tau_{x_i}(V_{\text{step},1})}], \label{eq:VC-analysis-x-sub-3ts-2}
\end{align}
which is simply the solution of the gating variable equations at $t=3\tau_s+3\tau_f$ for the voltage clamp experiment shown in Figure~\ref{fig:VC-step-3ts}. The limits for the time constants going to zero and infinity are similar to those in the case of two-timescales:
\begin{align}
    \lim_{\tau_{x_i} \to 0} x_i &= x_{i,\infty}(V_{\text{step},2})\\
    \lim_{\tau_{x_i} \to +\infty} x_i &= x_{i,\infty}(V_0),
\end{align}
while $x_i \approx x_{i,\infty}(V_{\text{step},1})$ if $\tau_{x_i} \approx \tau_s$ and $\tau_s \gg \tau_f$.

\subsection{Pre-compensation of voltages in absence of timescale separation}
\label{sec:VC-analysis-ident-i-ion-1-comp}

The method in the previous sections assumes that each gating variable of the biophysical model can be grouped into one of three categories: fast, slow, or ultraslow. Furthermore, the approximations rely on those timescales to be well separated from each other.

When the model does not show clear timescale separation, the assumption that the slower voltages will not significantly change after a step does not hold anymore. As long as the dynamics of the different gating variables can still be grouped in two or three timescales, it is possible to modify the described methods to compensate for the dynamics of the slower variables during the step experiment. In essence, the voltages used in the voltage clamp step experiments are modified so that the slower voltage(s) reach the desired values at the specified times $t=3\tau_f$ (or $t=3\tau_f+3\tau_s$ for three-timescale models) the slower voltage(s) have reached the desired values. This is illustrated in Figure~\ref{fig:VC-step-comp}.

In the case of two timescales, only the initial voltage of the voltage clamp step needs to be changed. The value of $V_s$ at $t=3\tau_f$ can be computed given the initial voltage $V_0$ and the step voltage $V_{\text{step}}$:
\begin{align}
    V_s(3\tau_f) &= V_0 + (V_{\text{step}}-V_0) \left(1-e^{-3\tau_f/\tau_s}\right).
\end{align}
For this voltage to reach the desired value $V_s^*$ at $t=3\tau_f$, the above expression can be solved for $V_0$:
\begin{align}
    V_0 &= \left[V_s^* - V_{\text{step}} \left(1-e^{-3\tau_f/\tau_s}\right)\right] / e^{-3\tau_f/\tau_s}.
\end{align}
Making this additional substitution in \eqref{eq:VC-analysis-x-sub-2ts} results in the substitution expression to obtain the value for $I_{\text{ion}}(V_{\text{step}},V_s^*)$.

\includetikzfigure{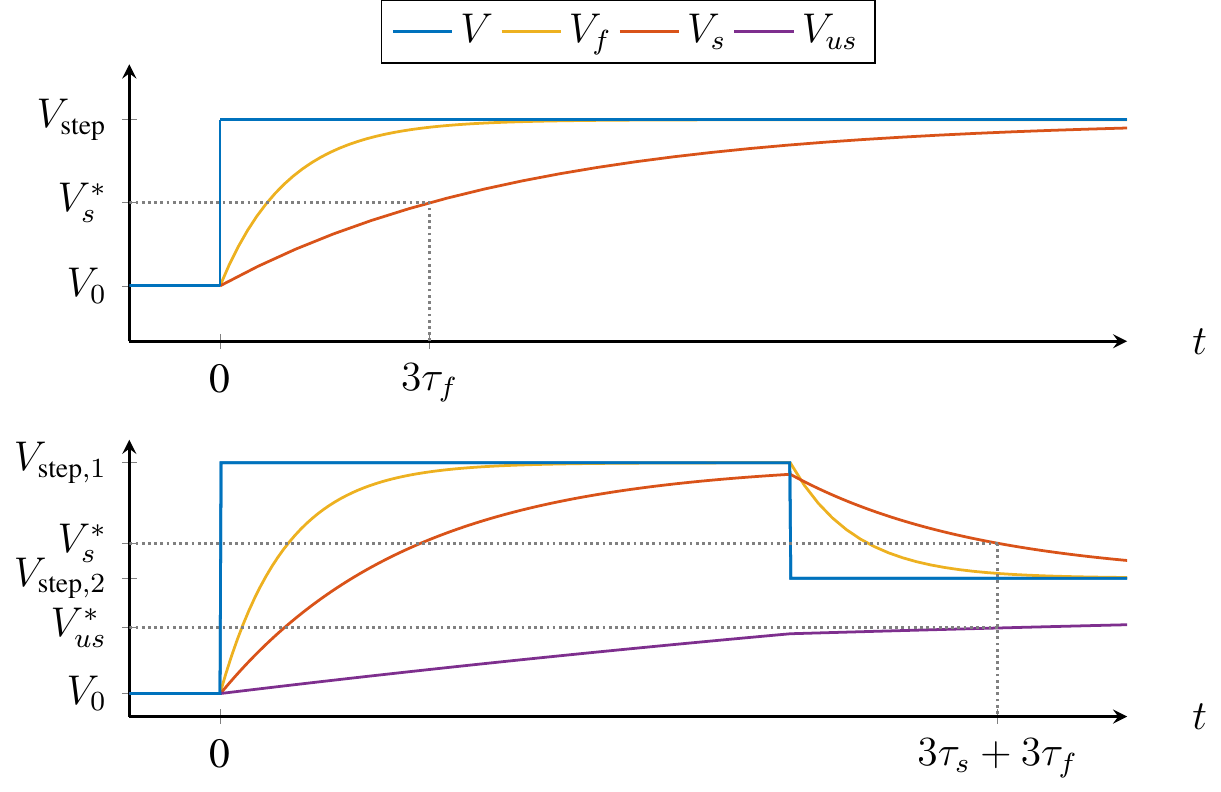}{.8}{.3}{Illustration of the pre-compensation of step voltages to reach the desired values of the slow voltages at $t=3\tau_f$ or $t=3\tau_f+3\tau_s$. Top: the choice of $V_0$ makes $V_s=V_s^*$ at $t=3\tau_f$ for two-timescale models. Bottom: the choice of $V_0$, $V_{\text{step},1}$ and  $V_{\text{step},2}$ leads to $V_s=V_s^*$ and $V_{us}=V_{us}^*$ at $t=3\tau_f+3\tau_s$ for three-timescale models.}

The same idea can be applied to three-timescale models. $V_{\text{step},1}$ is calculated in the same way as $V_0$ in the two-timescale case:
\begin{align}
    V_{\text{step},1} &= \left[V_s^* - V_{\text{step},2} \left(1-e^{-3\tau_f/\tau_s}\right)\right] / e^{-3\tau_f/\tau_s}.
\end{align}
This can be done because $V_s$ can be assumed to be at steady state at $t=3\tau_s$. $V_0$ can then be computed using $V_{\text{step},1}$. For simplicity, it is assumed that $\tau_{us} \gg \tau_f$, giving:
\begin{align}
    V_0 &= \left[V_{us}^* - V_{\text{step},1} \left(1-e^{-3\tau_s/\tau_{us}}\right)\right] / e^{-3\tau_s/\tau_{us}}.
\end{align}
If $\tau_{us} \gg \tau_f$ does not hold, the expression becomes slightly more complicated:
\begin{align}
    V_0 &= \left[\bar{V} - V_{\text{step},1} \left(1-e^{-3\tau_s/\tau_{us}}\right)\right] / e^{-3\tau_s/\tau_{us}}\\
    \bar{V} &= \left[V_{us}^* - V_{\text{step},2} \left(1-e^{-3\tau_f/\tau_{us}}\right)\right] / e^{-3\tau_f/\tau_{us}}.
\end{align}

After making these substitutions in \eqref{eq:VC-analysis-x-sub-3ts-1}-\eqref{eq:VC-analysis-x-sub-3ts-2}, the substitution expressions can be used to obtain the value for $I_{\text{ion}}(V_{\text{step},2},V_s^*,V_{us}^*)$.

\subsection{Calcium dynamics}
\label{sec:VC-analysis-ident-i-ion-1-ca}

Many biophysical models also contain calcium-gated ion channels apart from the more common voltage-gated ion channels. The conductance of these channels is usually described by a (nonlinear) function of the calcium concentration instead of a product of gating variables. The calcium dynamics obey a system of differential equations of the type
\begin{align}
    \tau_{[\text{Ca}^{\text{2+}}]} \dot{[\text{Ca}^{\text{2+}}]} &= [\text{Ca}^{\text{2+}}]_{\infty}(V, x_{\text{Ca},i}, \ldots) - [\text{Ca}^{\text{2+}}]\\
    \tau_{x_{\text{Ca},i}}(V) \dot{x}_{\text{Ca},i} &= x_{\text{Ca},i,\infty}(V) - x_{\text{Ca},i}\\
    & \ldots \nonumber
\end{align}
where $x_{\text{Ca},i}$ are the gating variables of the calcium channels. This will generally not result in a simple substitution expression. However, under the assumption that the calcium concentration changes much slower that the calcium gating variables, an approximate expression can be obtained. Assuming the calcium gating variables are at steady state whenever the voltage is constant during the voltage-clamp experiment, the expressions of the previous sections can be reused. $[\text{Ca}^{\text{2+}}](V_s)$ and $[\text{Ca}^{\text{2+}}](V)$ in these expressions are obtained by substituting the values of the calcium gating variables at $t=3\tau_s$ and $t=3\tau_s+3\tau_f$ respectively. Because of its simplicity, this approximation will be used in the rest of this paper.

\section{Iterative optimisation of the structural parameters}
\label{sec:VC-analysis-init-opt}

The previous section showed how to identify the function $I_{\text{ion}}$ for a given set of structural parameters. In this section we discuss how to initialise those parameters and how to iteratively optimise them using current clamp data.

\subsection{Initialisation of the time constants}
\label{sec:VC-analysis-init-time-constants}

To estimate the time constants, we assume that the biophysical model can be described by the multi-scale model \eqref{eq:VC-analysis-reduced-model-1}--\eqref{eq:VC-analysis-reduced-model-3} in the subthreshold regime. This is of course an approximation, but allows to find a simple procedure to find an initialisation of the time constants.

The Wiener structure of parallel linear systems followed by a static nonlinearity (Figure~\ref{fig:Wiener-model}) can be exploited to estimate the time constants of the model. Applying a sufficiently small voltage clamp signal will reveal the dynamics of the linear systems, since
\begin{align*}
    I_{\text{app}} &\approx C\dot{V} + \alpha_0 + \alpha_1 V + \alpha_2 V_f + \alpha_3 V_s + \alpha_4 V_{us}\\
    \alpha_1 &= \frac{\partial{I_{\text{ion}}}}{\partial{V}}, \alpha_2 = \frac{\partial{I_{\text{ion}}}}{\partial{V_f}}, \ldots
\end{align*}
where the coefficients $\alpha_i$ depend on the membrane potential around which the experiment is performed. The contribution of $C\dot{V}$ is easily removed as it only produces a sharp pulse at the time of the step. Repeating this experiment at multiple subthreshold voltages gives a more robust estimation, as some of the coefficients $\alpha_i$ might be small around a single voltage.

Classical system identification techniques can then be used to obtain the parameters of the linear systems from this voltage clamp data. We choose an extension by \citet{Miller2013} of the Ho-Kalman-Kung algorithm \citep{Ho1966,Zeiger1974,Kung1978} for step responses. The measured voltage clamp step responses are treated as the different outputs of the system to a single step input. The algorithm allows to select a model order (number of voltage variables) based on the Hankel singular values. The poles can also be restricted to be real and stable, to fit the structure of the multi-scale model. The sought time constants $\tau_f, \tau_s, \tau_{us}$ will hence be the negative inverse of the identified poles.

The described method to obtain initial estimates for the time constants is by no means optimal or the only one possible. A common approach used in Wiener system identification is the ``best linear approximation'' (BLA) \citep{Enqvist2005a,Schoukens2012}. While this approach can result in more accurate estimates, it requires more data and the Gaussian input signals used are less suited to the voltage-clamp setting. Furthermore, the time constants only have to be accurate enough to make the optimisation procedure of Section~\ref{sec:VC-analysis-param-opt} converge to a reasonable set of parameters.

\subsection{Initialisation of the other structural parameters}
\label{sec:VC-analysis-init-params}

Apart from the time constants and $I_{\text{ion}}$, the only parameters left to identify in \eqref{eq:VC-analysis-NLIF-1}--\eqref{eq:VC-analysis-NLIF-3} are the membrane capacitance $C$, the reset parameters $V_r$, $V_{s,r}$ and $\Delta V_{us}$, and the cutoff voltage $V_{\text{max}}$. Those parameters are easily estimated from current clamp simulations of the biophysical model.

$V_{\text{max}}$ is taken to be just after the onset of the spike, around the maximum of the first derivative of the membrane potential. As long as $V_{\text{max}}$ lies sufficiently above threshold, its precise value should not influence the behaviour of the integrate-and-fire model. It is therefore fixed during the optimisation step.

The voltage reset $V_r$ is always taken to be equal to $V_{\text{max}}$. The reset values of the slower voltages are parameters to be optimised, but are physiologically constrained. A reasonable initialisation is to set $V_{s,r}$ sufficiently (e.g. \SI{20}{\mV}) above $V_r$. The parameter $\Delta V_{us}$ is constrained to be positive, since $V_{us}$ would only increase after a spike in the model \eqref{eq:VC-analysis-reduced-model-1}--\eqref{eq:VC-analysis-reduced-model-3}. Therefore it can be initialised at \SI{0}{\mV}.

Finally, the membrane capacitance $C$ is initially set to \SI{1}{\micro\F\per\cm\squared}, as is often the case in a conductance based model.

\subsection{Local parameter optimisation}
\label{sec:VC-analysis-param-opt}

The structural parameters of the model can be iteratively optimised to improve the matching of representative current clamp data (see e.g. Figure~\ref{fig:CS-test-data-gA-0}, top).

We choose to minimise a cost function based on the residual current, which has been used before for the parameter estimation of biophysical models \citep{Morse2001,Huys2006,Lepora2012}. The residual current for our model is defined as
\begin{equation}
    \label{eq:VC-analysis-cost-fun}
    I_\text{res} = C\dot{V} - I_{\text{app}} + I_{\text{ion}}(V,V_s,V_{us})
\end{equation}
and is zero when the membrane current equation is satisfied. The test data provides $V$ and $I_{\text{app}}$, while the derivative $\dot{V}$ can be obtained by numeric differentiation. The term $I_{\text{ion}}$ depends on the chosen time constants and to evaluate it, the voltage trace is filtered by the corresponding first-order linear low-pass filters. The action potentials themselves are eliminated by removing the data for which $V > V_{\text{max}}$, as the model is only optimised in the subthreshold regime. Instead, the different voltages are reset after every spike using their respective reset parameters, as shown in Figure~\ref{fig:CS-test-data-gA-0} (bottom). The voltage traces of Figure~\ref{fig:CS-test-data-gA-0} (bottom) can then be used to evaluate $I_\text{ion}(V,V_s,V_{us})$ and thus the residual current.

\includetikzfigure{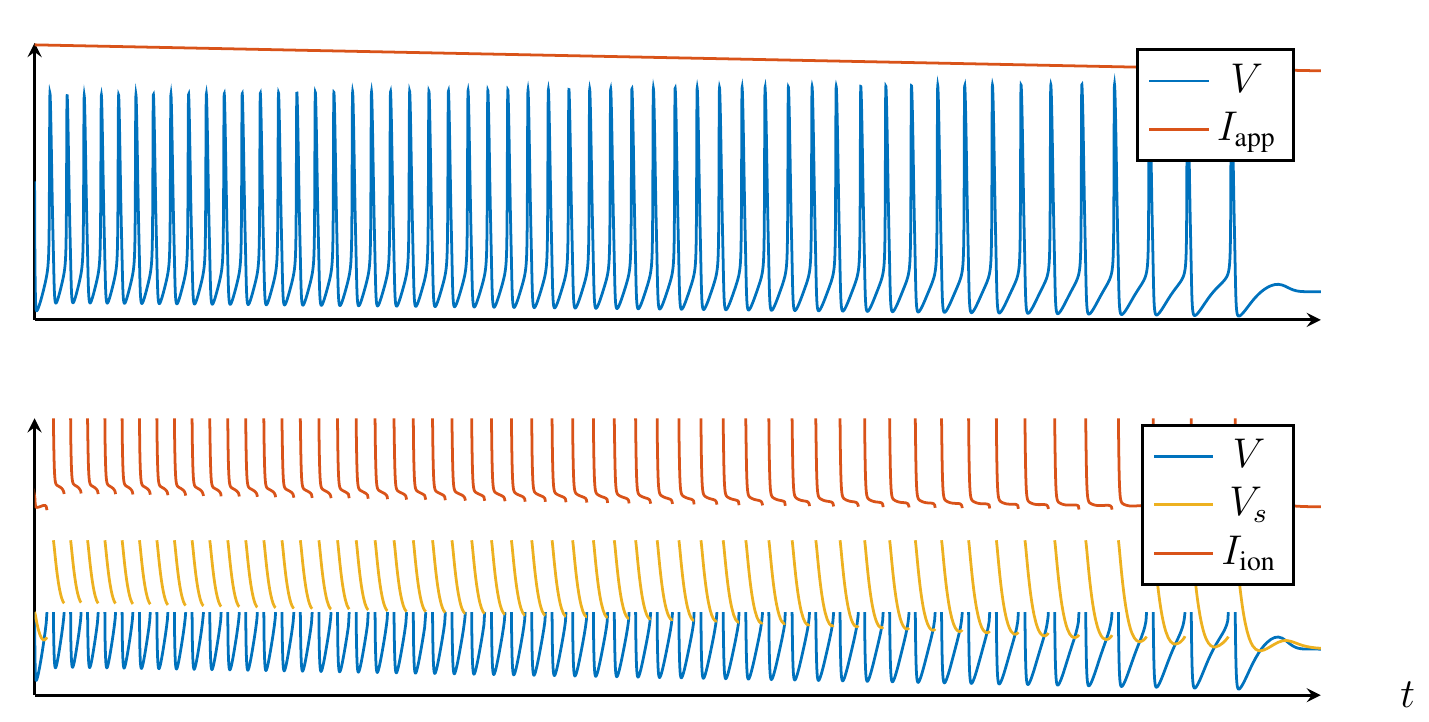}{1}{.3}{Test data generated using the Connor-Stevens model with $\bar{g}_A=\SI{0}{\milli\siemens\per\cm\squared}$. Top: Current clamp experiment with a decreasing ramp of the current $I_\text{app}$. Bottom: Data used to evaluate the cost function, where points for which $V>V_{\text{max}}$ are removed. $V_s$ is obtained by filtering $V$ with a first-order linear low-pass filter and reset to $V_{s,r}$ after every spike, and $I_\text{ion}$ is evaluated using these $V$ and $V_s$.}

It was found that simple least-squares minimisation was sufficient for the purpose of this paper. Apart from the time constants, the free parameters in the optimisation are $C$ and the reset values $V_{s,r}$ and $\Delta V_{us}$. The advantage of using the residual current over the residual voltage for the cost function is that the former does not suffer from the extreme sensitivity to tiny variations in the spike timing. This can be observed in Figure~\ref{fig:CS-tau-opt-cost}: the cost function based on the residual current is locally convex. In contrast, a cost function based on the residual voltage has many local minima and is much more irregular.

\includetikzfigure{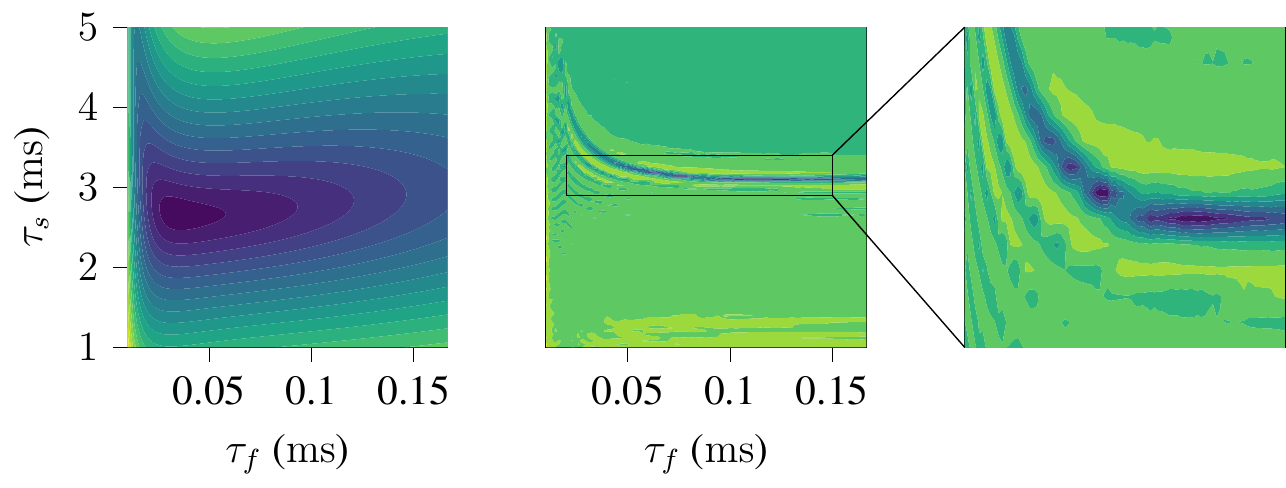}{.33}{.33}{Contour plots of two least squares cost functions as a function of $\tau_f$ and $\tau_s$ for the Connor-Stevens model with $\bar{g}_A=\SI{0}{\milli\siemens\per\cm\squared}$. Left: cost function based on the residual current, as used in this paper. Centre \& right: cost function based on the residual voltage ($V_\text{res}=V-V^*$, where $V$ is the voltage of the test data and $V^*$ is the voltage of the simulation of the integrate-and-fire model).}

An important property of the multi-scale integrate-and-fire model is that its structural parameters have a clear interpretation. This makes it easy to initialise them at a reasonable value by inspection of a few voltage and current clamp experiments. The additional optimisation procedure is a straightforward least-squares local optimisation.

\section{Integrate-and-fire modelling and phase portrait analysis}
\label{sec:VC-analysis-PP-analysis}

The benefit of the proposed integrate-and-fire model is not purely computational. We now show that it is also amenable to phase portrait analysis, which provides mathematical insight on the initial biophysical model, regardless of its dimension.

For a two-timescale analysis, the phase portrait of the integrate-and-fire model is entirely characterised by two curves: the $V$-nullcline $I_{\text{ion}}(V,V_s)=I_{\text{app}}$, and the $V_s$-nullcline $V_s=V$. In other words, the level curves of the identified ion current $I_{\text{ion}}$ determine the phase portrait of the model.

When the integrate-and-fire model is three-dimensional, we can describe its dynamics by considering the ultraslow variable $V_{us}$ as a bifurcation parameter and by studying the family of phase portraits parameterised by $V_{us}$. In that sense, it can be said that the proposed integrate-and-fire model maps an arbitrary biophysical model to a family of phase portraits. This is very convenient for a qualitative understanding of the dynamical properties of the model.

The following sections will illustrate the identification procedure and the phase portrait analysis of the integrate-and-fire model. The model is identified on two biophysical models from the literature. The first is the classical Connor-Stevens model, whose behaviour can be easily modulated by a change of maximal conductance. The second is the model by \citet{Drion2018}, exhibiting a switchable slow negative conductance.

\subsection{Modulation of excitability type in the Connor-Stevens model}
\label{sec:VC-analysis-modulation-excitability}

The Connor-Stevens model \citep{Connor1971,Connor1977} is a six-dimensional conductance-based model for gastropod neuron somas. It has all the variables of the Hodgkin-Huxley model in addition to an extra potassium current $I_A$. One of its characteristic features is Type~I excitability: the spiking frequency approaches zero when the applied current approaches the rheobase. This is in contrast to the Type~II excitability of the Hodgkin-Huxley model, whose spiking frequency makes a jump as the applied current is increased.

\subsubsection{A two-timescale integrate-and-fire approximation of the Connor-Stevens model}

Considering that this model was the example used by \citet{Kepler1992} for the equivalent potentials method, the method of this paper is expected to work well on this model. Similar to the work of Kepler et al., we start by a two-dimensional reduction of the model using the method of Section~\ref{sec:VC-analysis-ident-i-ion-1-2ts}. The time constants found after the optimisation procedure are $\tau_f=\SI{0.022}{\ms}$ and $\tau_s=\SI{6.7}{\ms}$. They reflect the timescale separation of the gating variables, which span a range of \SIrange{0.03}{3}{\ms}.

The phase portrait of the two-timescale continuous-time model is displayed in Figure~\ref{fig:CS-VC-PP-2} (left). For a specific value of the applied current $I^*_\text{app}$, the $V_s$-nullcline intersects the $V$-nullcline at the transcritical singularity \citep{Franci2012}. As the current increases from below to above the value $I^*_\text{app}$, stability is lost in a saddle-node on invariant circle (SNIC). This was shown to be the mechanism of Type~I excitability in this model in \citet{Drion2015a}. The phase portrait of Figure~\ref{fig:CS-VC-PP-2} (right) shows the trajectory of spiking in the integrate-and-fire model \eqref{eq:VC-analysis-NLIF-1}--\eqref{eq:VC-analysis-NLIF-3} in orange. Since the model has a reset mechanism to replace the action potential generation, the periodic spiking occurs due to a hybrid limit cycle.

\includetikzfigure{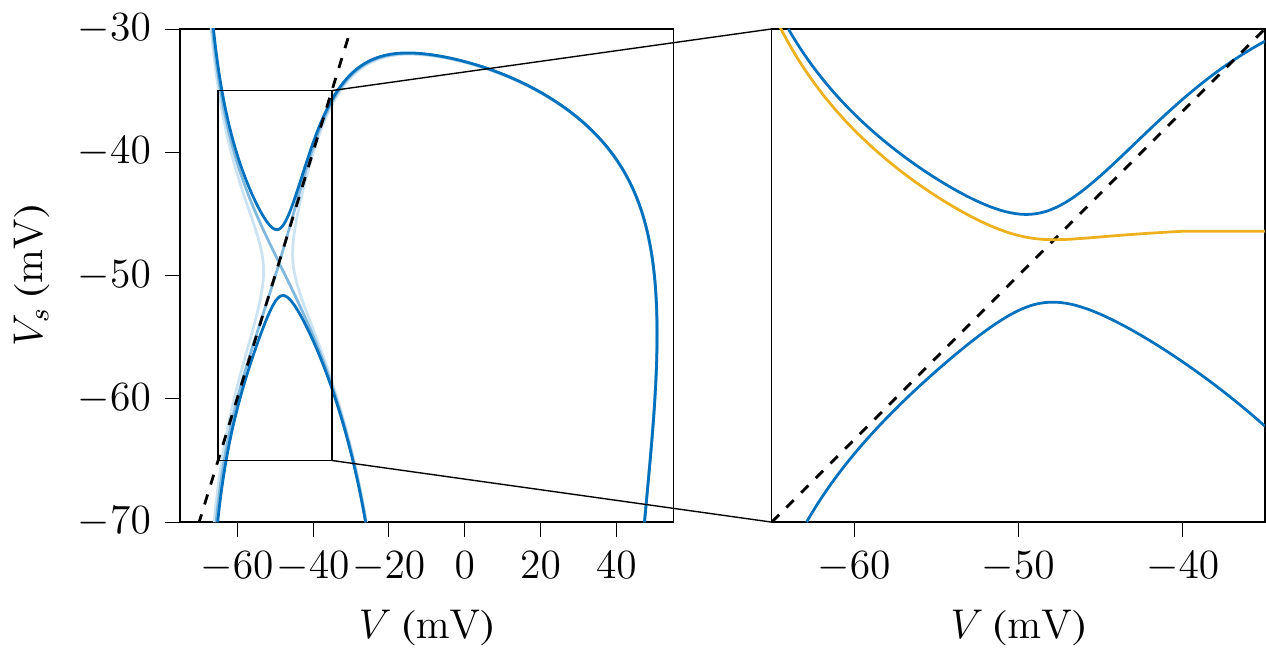}{.45}{.45}{Phase portraits of the integrate-and-fire model obtained from the Connor-Stevens model with standard parameters \citep{Connor1977}. The $V$- and $V_s$-nullclines are drawn as full and dashed lines respectively. Left: The $V$-nullclines are drawn for different values of $I_\text{app}$: low (light blue), $I_\text{app}$ at the transcritical bifurcation in fast subsystem (medium blue) and high (dark blue). Right: Detailed portion of the phase portrait (for the high value of $I_\text{app}$), together with the corresponding trace of the limit cycle oscillation in orange.}

Figure~\ref{fig:CS-VC-PP-gA} (top centre) illustrates the frequency-current (f-I) curves of the model. The bifurcation voltage is well predicted by the integrate-and-fire model. Although the onset of the f-I curve is sharper than for the original model, both curves eventually converge.

\includetikzfigure{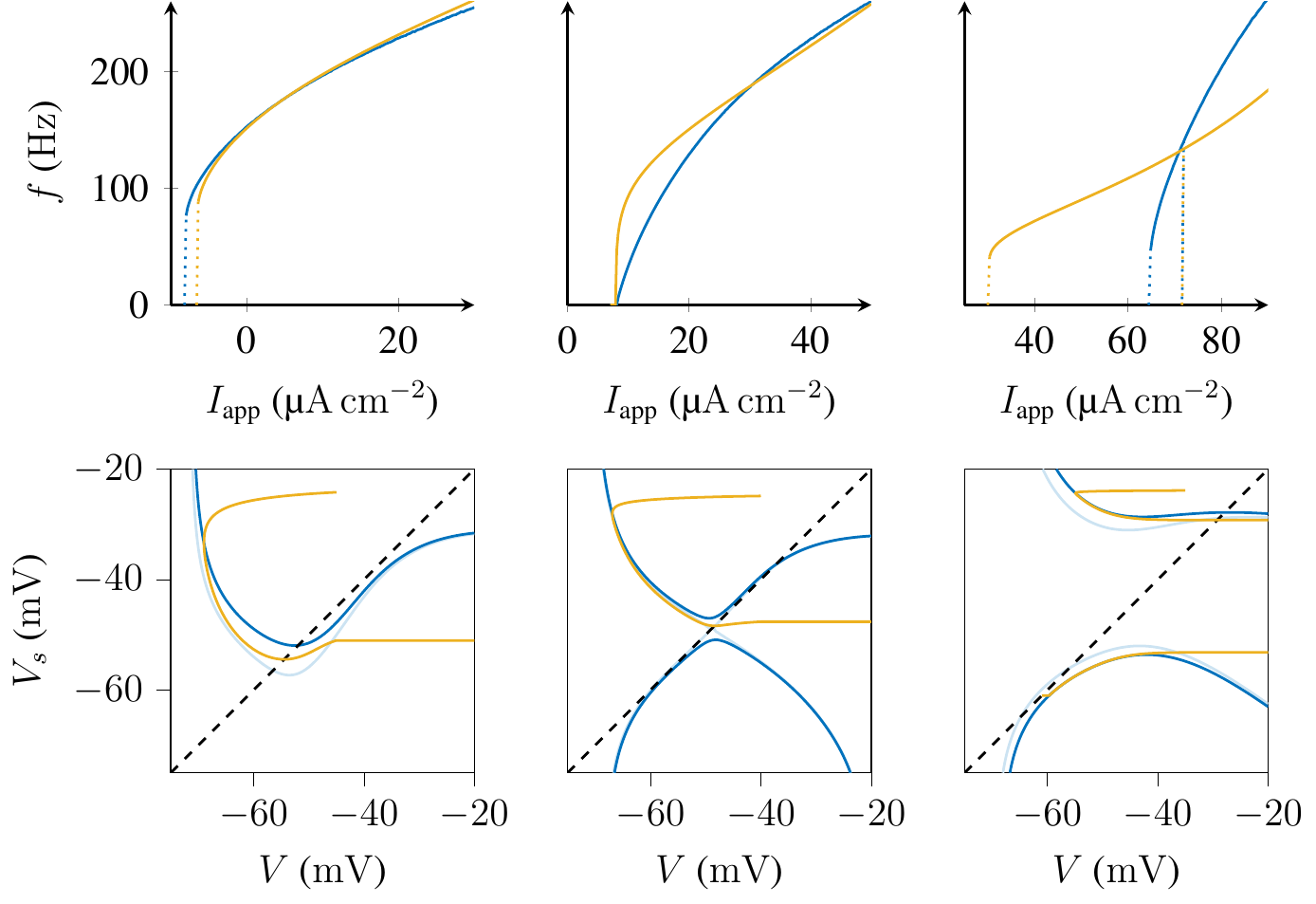}{.33}{.33}{Phase portraits of the integrate-and-fire model (bottom) and f-I curves (top) of the Connor-Stevens model (blue) and the integrate-and-fire model (orange) for different values of $\bar{g}_A$. The $V$- and $V_s$-nullclines are drawn as full and dashed lines respectively, the trajectories after losing stability in orange. Left: $\bar{g}_A=\SI{0}{\milli\siemens\per\cm\squared}$, centre: $\bar{g}_A=\SI{47.7}{\milli\siemens\per\cm\squared}$, right: $\bar{g}_A=\SI{200}{\milli\siemens\per\cm\squared}$.}

The excitability type of the model can be changed easily by modulating the maximal conductance $\bar{g}_A$. For $\bar{g}_A=\SI{0}{\milli\siemens\per\cm\squared}$, the model is similar to the Hodgkin-Huxley model, which exhibits Type~II excitability. We again obtain a two-timescale integrate-and-fire approximation of this model by finding $I_\text{ion}$ and the structural parameters using the new value for $\bar{g}_A$. Its phase portrait is shown in Figure~\ref{fig:CS-VC-PP-gA}~(bottom left) and is qualitatively the same as that of the FitzHugh-Nagumo model~\citep{FitzHugh1961,Nagumo1962}. There is a subcritical Hopf bifurcation resulting in Type~II excitability, confirmed by the f-I curve in Figure~\ref{fig:CS-VC-PP-gA}~(top left). The integrate-and-fire model loses stability at a slightly higher value of $I_\text{app}$ than the original model, but both f-I curves remain close to each other afterwards.

On the other hand, increasing $\bar{g}_A$ above its nominal value of \SI{47.7}{\milli\siemens\per\cm\squared} to \SI{200}{\milli\siemens\per\cm\squared} results in a bistable phase portrait (Figure~\ref{fig:CS-VC-PP-gA}, bottom right): for a specific range of $I_\text{app}$ a stable hybrid limit cycle on the upper branch coexists with a stable fixed point on the lower branch of the $V$-nullcline. The fixed point loses stability in a saddle-node bifurcation, while the limit cycle disappears in a fold limit cycle bifurcation. This was called Type~II* excitability in \citet{Drion2015a}, which is similar to Type~II excitability, but has a hysteretic f-I curve. Figure~\ref{fig:CS-VC-PP-gA} (top right) shows that the integrate-and-fire model captures this hysteresis. Although the model starts spiking at a lower value of $I_\text{app}$, the saddle-node bifurcation occurs at the same point. This can be explained by the fact that the fold limit cycle bifurcation is a global bifurcation, which is harder to capture than the local saddle-node bifurcation.

\subsubsection{A three-timescale integrate-and-fire approximation of the Connor-Stevens model}

The results in the previous section show that the two-timescale reduction of the Connor-Stevens model neuron model is useful to obtain a qualitative approximation of the behaviour of the original model. The modulation of excitability type can be predicted from the phase portraits and it is possible to construct an integrate-and-fire model that replicates this modulation. However, the resulting integrate-and-fire model lacks a quantitative approximation of the voltage trace (not shown) and f-I curve. Kepler et al. improved the quality of their reduction of the Connor-Stevens model by adding a third equivalent potential.

In an analogous effort, we approximate the original model by a three-timescale integrate-and-fire model using the method of Section~\ref{sec:VC-analysis-ident-i-ion-1-3ts}. The three timescales are not strongly separated, however, and the method of Section~\ref{sec:VC-analysis-ident-i-ion-1-comp} was used to adjust for this lack of separation. This resulted in a model with the time constants $\tau_f=\SI{0.037}{\ms}$, $\tau_s=\SI{1.7}{\ms}$ and $\tau_{us}=\SI{2.8}{\ms}$. It is hypothesised that, like in the equivalent potentials method, this third timescale is necessary to account for the dynamics of the inactivation variable of the current $I_A$ which are different from those of the other gating variables. Figure~\ref{fig:CS-VC-3ts} (right) shows that the f-I curve for this three-timescale model almost perfectly match that of the original model. A comparison of a current clamp simulation with a linearly increasing applied current for both models (Figure~\ref{fig:CS-VC-3ts}, left) reveals very similar voltage traces.

\includetikzfigure{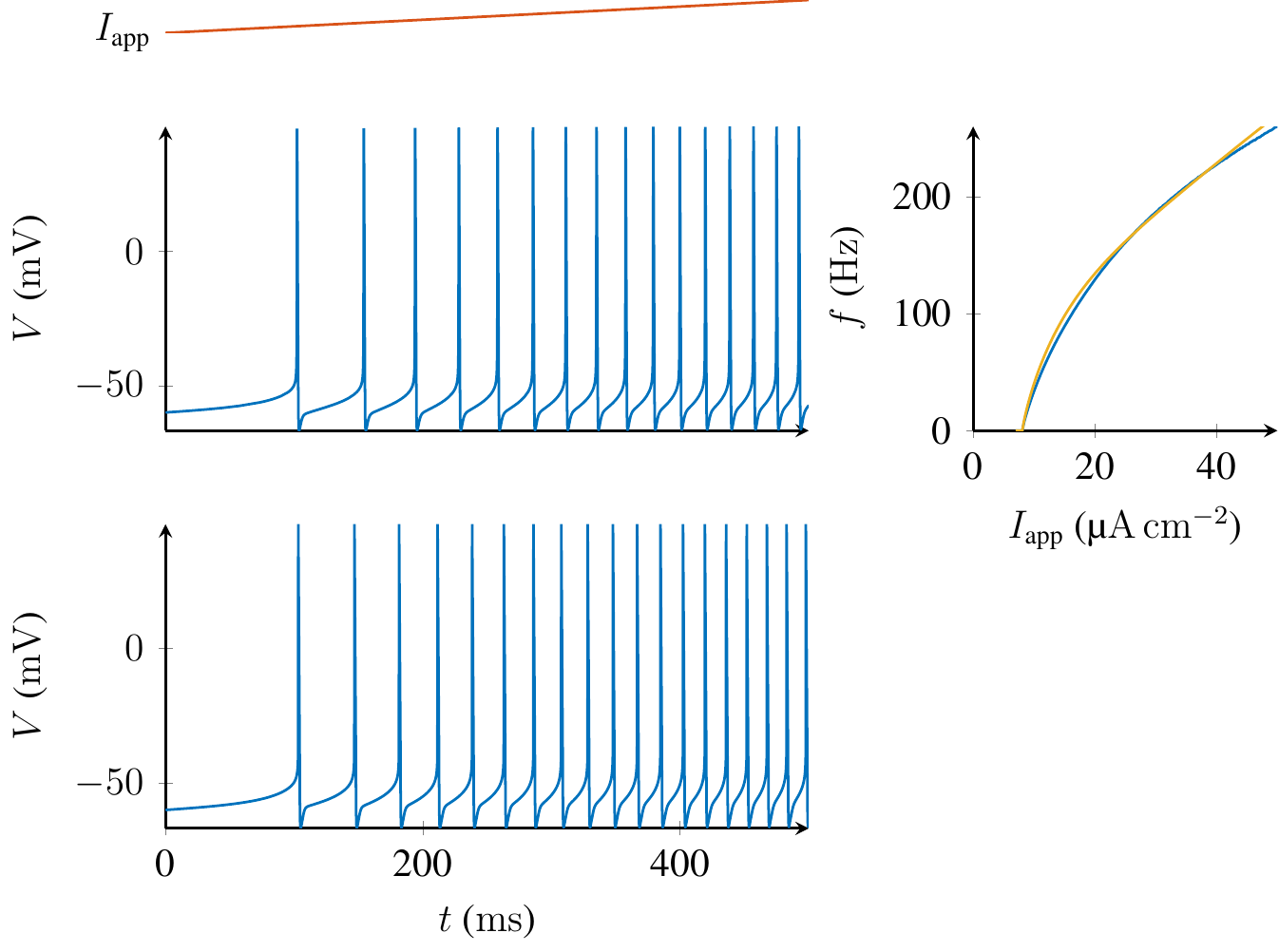}{.33}{.33}{Left: voltage traces of the original Connor-Stevens model (top) and the three-timescale integrate-and-fire model (bottom) for $I_\text{app}$ linearly increasing from $8$ to $\SI{12}{\uA\per\square\cm}$. Right: f-I curves for the original Connor-Stevens model (blue) and the three-timescale integrate-and-fire model (orange).}

While the f-I curves match well for the three-timescale model with $\bar{g}_A=\SI{47.7}{\milli\siemens\per\cm\squared}$, we did not find a similar improvement for $\bar{g}_A=\SI{200}{\milli\siemens\per\cm\squared}$ (not shown). More work is necessary to determine whether the method is able to find a better approximation by using different current clamp data for the optimisation of the structural parameters.

\subsection{Modulation between spiking and bursting due to a switchable slow negative conductance}
\label{sec:VC-analysis-modulation-spiking-bursting}

Our second illustration uses the eight-dimensional conductance-based model introduced in \citet{Drion2018}. A critical physiological feature of this model is a T-type calcium channel with low-threshold activation in the slow timescale. T-type calcium channels endow the model with slow regenerativity by having an activation which is slower than the sodium channel activation. Furthermore, the channels are inactivated at a relatively low threshold (on an even slower timescale). This results in a slow negative conductance that is switchable by an external current: it is only switched on by hyperpolarisation. This current was hypothesised in \citet{Drion2018} as a critical mechanism for the neuromodulation of network states. Figure~\ref{fig:Drion-VC-step}~(left) shows how the switchable negative conductance can be observed from a voltage clamp step experiment. The slope of the current response in the slow timescale determines the absence or presence of slow regenerativity~\citep{Franci2017}.

\includetikzfigure{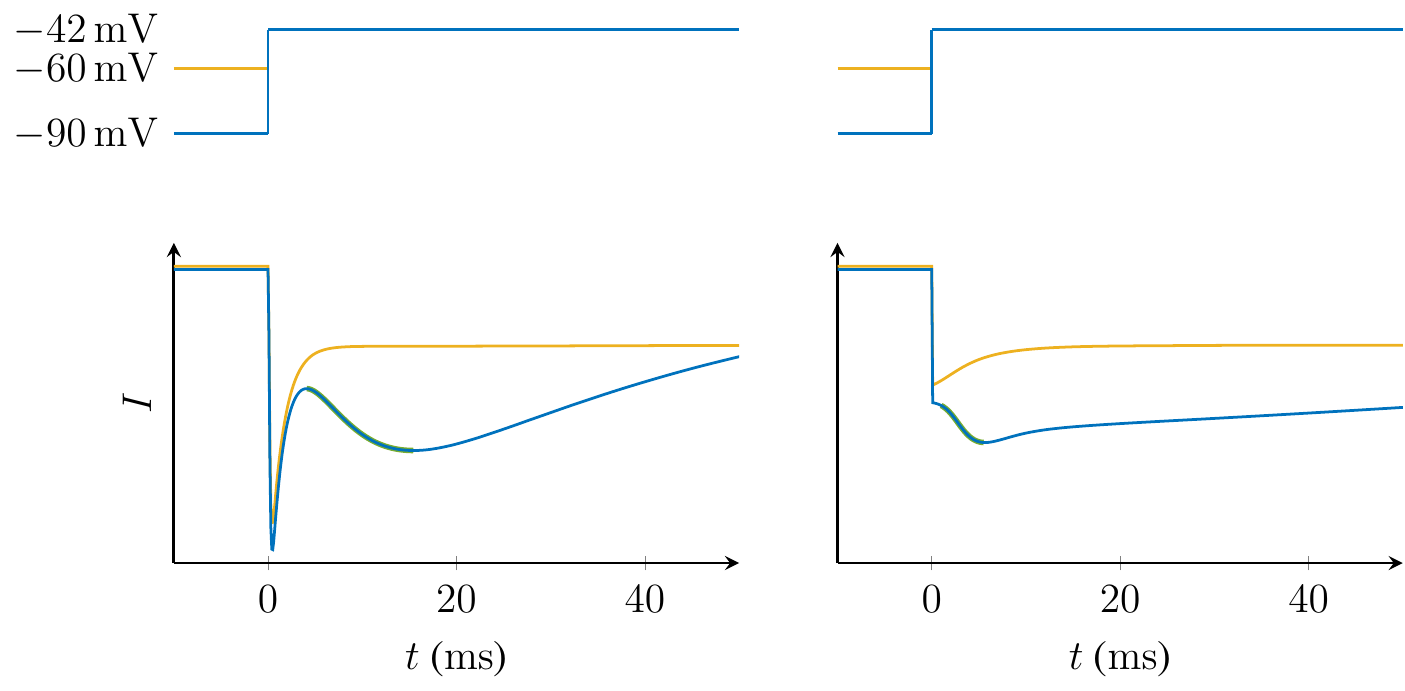}{.5}{.33}{Comparison of the response to a voltage clamp step from a hyperpolarised (blue) and depolarised (orange) state for the original model of~\citet{Drion2018} (left) and its reduction (right). The voltage is stepped from \SI{-90}{\mV} and \SI{-60}{\mV} to \SI{-42}{\mV}. The part of the responses highlighted in green shows the presence of a slow negative conductance in the hyperpolarised state and its absence in the depolarised state.}

A distinctive behaviour of the model in current clamp is hyperpolarisation-induced bursting (HIB). This means the model can be switched from slow spiking to bursting by sufficiently lowering the input current, as shown in Figure~\ref{fig:Drion-hybrid}~(top).

We use the method of Section~\ref{sec:VC-analysis-ident-i-ion-1-3ts} to construct an integrate-and-fire approximation of the biophysical model of \citet{Drion2018}. The calcium dynamics are treated as explained in Section~\ref{sec:VC-analysis-ident-i-ion-1-ca}. Its behaviour can be studied using three timescales. To visualise the obtained reduced model, phase portraits in the $V$-$V_s$ space are shown for different values of $V_{us}$. Assuming the ultraslow timescale is much slower than the slow timescale, the behaviour of the model can be analysed by looking at the fast-slow system in these phase portraits.

For $I_\text{app}=\SI{0}{\uA\per\square\cm}$, the original model shows regular spiking, with a lower spiking frequency than during the bursting (Figure~\ref{fig:Drion-spiking}, left). As this is a two-timescale behaviour, $V_{us}$ can be approximated by a constant value. Figure~\ref{fig:Drion-spiking} (right) shows the phase portrait corresponding to this situation. There is no bistability between the resting and spiking state, and the phase portrait is the standard phase portrait of a spiking model.

\includetikzfigure{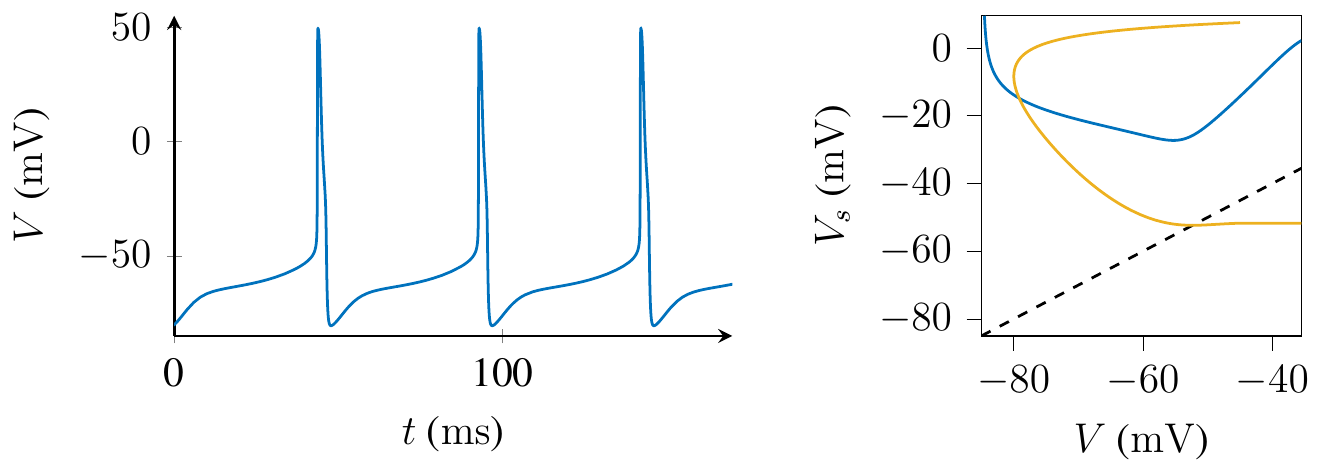}{.33}{.33}{Left: Voltage trace of the model of \citet{Drion2018} for $I_\text{app}=\SI{0}{\uA\per\square\cm}$. Right: Phase portrait of the three-dimensional integrate-and-fire approximation showing the absence of rest-spike bistability. The $V$- and $V_s$-nullclines are drawn as full and dashed lines respectively, the stable (hybrid) limit cycle in orange.}

When the current is lowered to $I_\text{app}=\SI{-1.6}{\uA\per\square\cm}$, the model bursts (Figure~\ref{fig:Drion-bursting}, top). The phase portraits during the different phases of the burst are shown in Figure~\ref{fig:Drion-bursting} (bottom). The burst is initiated by the loss of stability in a saddle-node bifurcation on the lower branch of the $V$-nullcline (Figure~\ref{fig:Drion-bursting}, bottom left). The subsequent spiking, on the limit cycle on the upper branch of the $V$-nullcline, causes $V_{us}$ to increase. This increase moves the two branches of the $V$-nullcline closer, resulting in bistability between a limit cycle and a stable fixed point (Figure~\ref{fig:Drion-bursting}, bottom centre). As $V_{us}$ increases even further, the limit cycle is lost in a saddle-homoclinic bifurcation, moving the trajectory to the stable fixed point and thus terminating the burst (Figure~\ref{fig:Drion-bursting}, bottom right).

\includetikzfigure{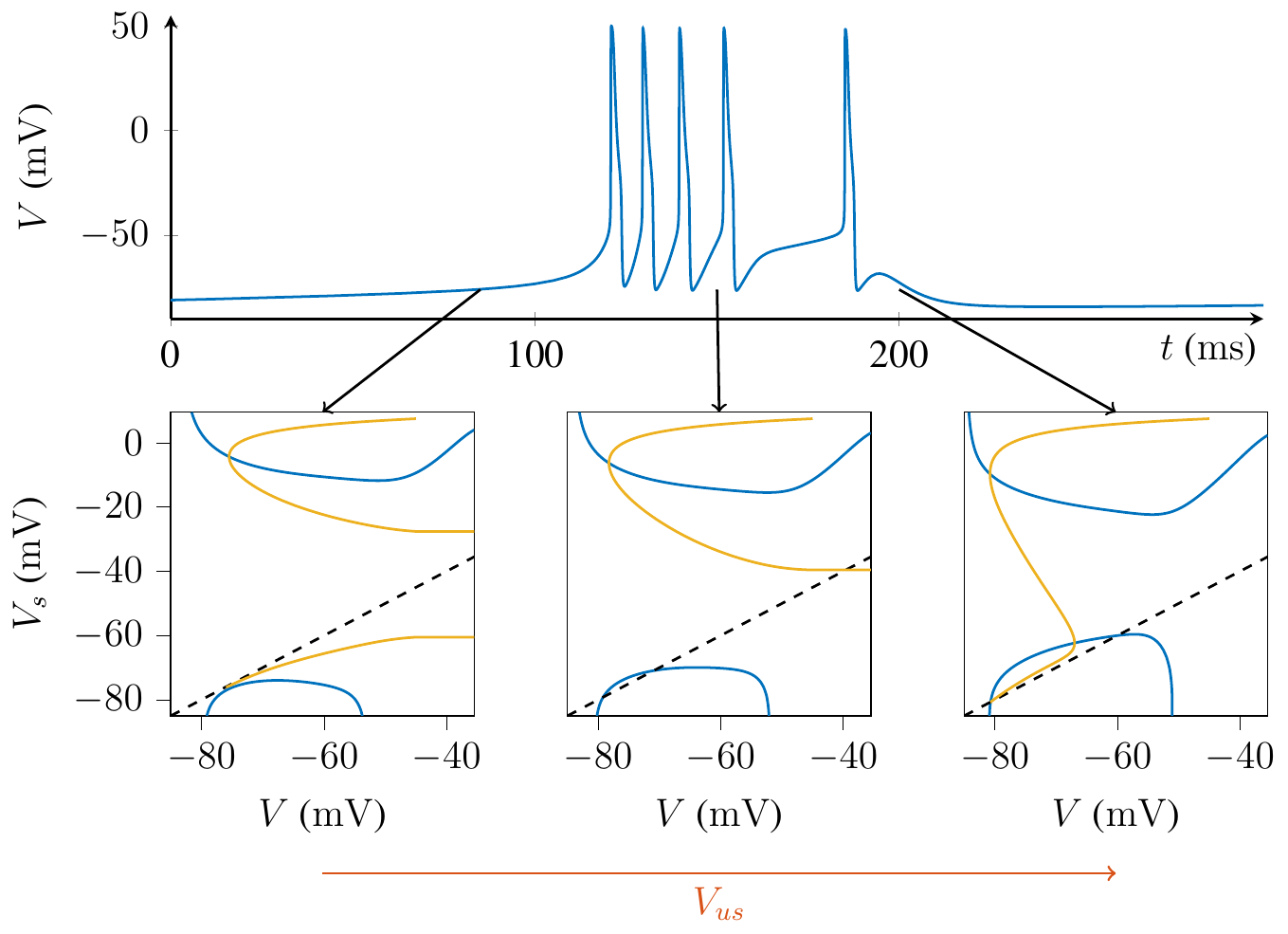}{.33}{.33}{Top: Voltage trace of a burst in the model of \citet{Drion2018} with $I_\text{app}=\SI{-1.6}{\uA\per\square\cm}$. Bottom: Portions of fast-slow phase portraits of the three-dimensional integrate-and-fire approximation for different values of $V_{us}$. The blue curve is the $V$-nullcline, the dashed curve is the $V_s$-nullcline, and each orange curve is a trajectory of the phase portrait that captures a piece of the bursting attractor in the three-dimensional model.}

Figure~\ref{fig:Drion-VC-step}~(right) shows the voltage clamp response of the reduced model for steps from two different voltages. Although different from the full biophysical model of Figure~\ref{fig:Drion-VC-step}~(left), the integrate-and-fire model retains the switchable negative conductance responsible for hyperpolarisation-induced bursting (HIB). Only for the hyperpolarised step, the slow response has a negative slope (highlighted in green). The fast response is instantaneous because $V_f$ and $V$ are a single variable in the reduced model.

Again, the obtained reduction can be used to construct an integrate-and-fire model. Figure~\ref{fig:Drion-hybrid} shows the voltage trace before and after a hyperpolarising step current for the original model (top) and the reduced model (bottom). While the result is not a perfect quantitative match, it is quite accurate given the reduction for a reduction of the number of variables from eight to three.

\includetikzfigure{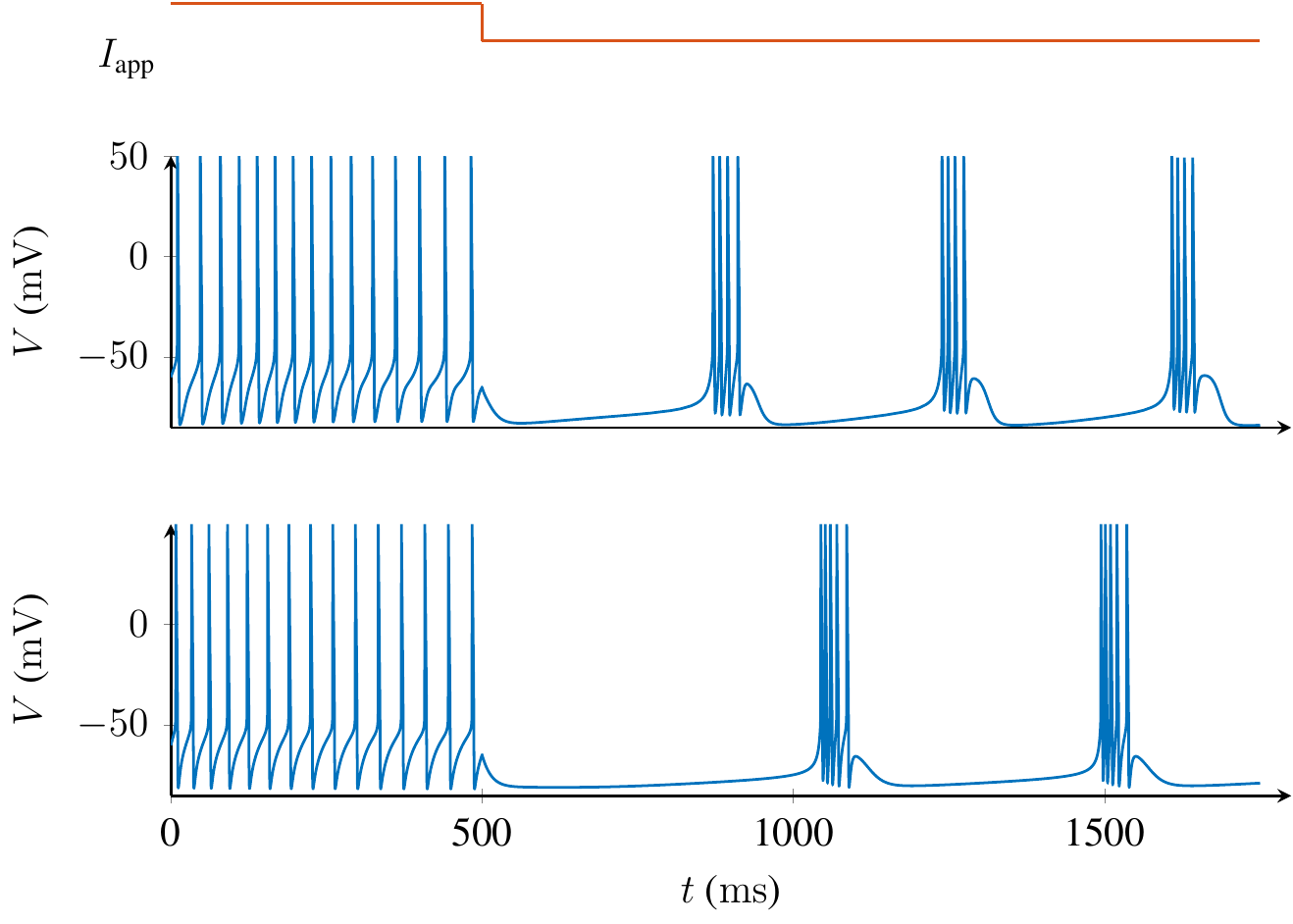}{.9}{.3}{Voltage trace of the model of \citet{Drion2018} (top) and its three-dimensional integrate-and-fire approximation (bottom) for $I_\text{app}$ stepping from 0 to \SI{-1.6}{\uA\per\square\cm} at $t=\SI{500}{\ms}$.}

\section{Discussion}
\label{sec:VC-analysis-discussion}

\subsection{Validity of the method}
\label{sec:VC-analysis-ident-i-ion-1-validity}

The method of this paper attempts to match the voltage clamp response of an integrate-and-fire and a biophysical model. The idea of modelling the voltage clamp response to obtain a description of the neural dynamics is exactly what Hodgkin and Huxley did in their seminal work \citep{Hodgkin1952}.

An important difference, however, is that we only match the response to a series of steps at a specific time. While this results in an analytical expression to reduce a biophysical model, it also means that the voltage clamp response of both models will not necessarily match for other inputs and at other times. Nevertheless, under certain assumptions (explained below) the proposed method can be seen as a reduction method similar to the method of equivalent potentials of \citet{Kepler1992}.

The notion of equivalent potentials is a way to convert the gating variables in a biophysical model into potentials or voltages. Every variable $x_i$ is simply replaced by the associated voltage of the steady-state function: $V_{x_i} = x^{-1}_{i,\infty}(x_i)$. In doing so, only the description of the system is changed, but not its input-output dynamics. The new form, however, can make it easier to discover relationships between different variables, necessary to reduce the system.

Kepler et al. propose such a reduction method by grouping equivalent potentials with similar dynamics and replacing them by a weighted average of their group. The weights are found by optimising the local approximation of the model. For the method to work, the dynamics of the equivalent potentials should fall into groups with similar dynamics. The method of this paper provides a simpler alternative based on voltage clamp experiments, by making the additional assumption that the dynamics of each group of slower equivalent potentials can be described sufficiently well by a first-order linear low-pass filter as in~\eqref{eq:VC-analysis-reduced-model-2}--\eqref{eq:VC-analysis-reduced-model-3}, at least in the subthreshold regime.

This assumption might not always fully hold in practice, but the method can then still produce an acceptable reduction, although this should be carefully validated afterwards.

\subsection{Local approximation by Multi-Quadratic Integrate-and-Fire model}
\label{sec:VC-analysis-approx-MQIF}

The Multi-Quadratic Integrate-and-Fire (MQIF) model~\citep{Drion2012,VanPottelbergh2018} is a special case of the multi-scale integrate-and-fire model \eqref{eq:VC-analysis-NLIF-1}--\eqref{eq:VC-analysis-NLIF-3}, in which $I_\text{ion}$ is a sum of quadratic functions in $V$, $V_s$ and $V_{us}$:
\begin{align}
    \label{eq:VC-analysis-MQIF}
    I_\text{ion}(V,V_s,V_{us}) &= \bar{g}_f(V-V^0)^2 + \bar{g}_s(V_s-V_s^0)^2 +\bar{g}_{us}(V_{us}-V_{us}^0)^2.
\end{align}
The quadratics in $V$ and $V_s$ provide a normal form of the transcritical singularity in the fast subsystem, organising the rest-spike bistability. The quadratic in $V_{us}$ models the ultraslow feedback necessary for bursting. The relative positions of the parameters $V^0$ and $V_s^0$ determine whether there is rest-spike bistability or not, but also influence the excitability type. This is illustrated in Figure~\ref{fig:MQIF-type}, which sketches the phase portraits for the three possible regimes.

\includetikzfigure{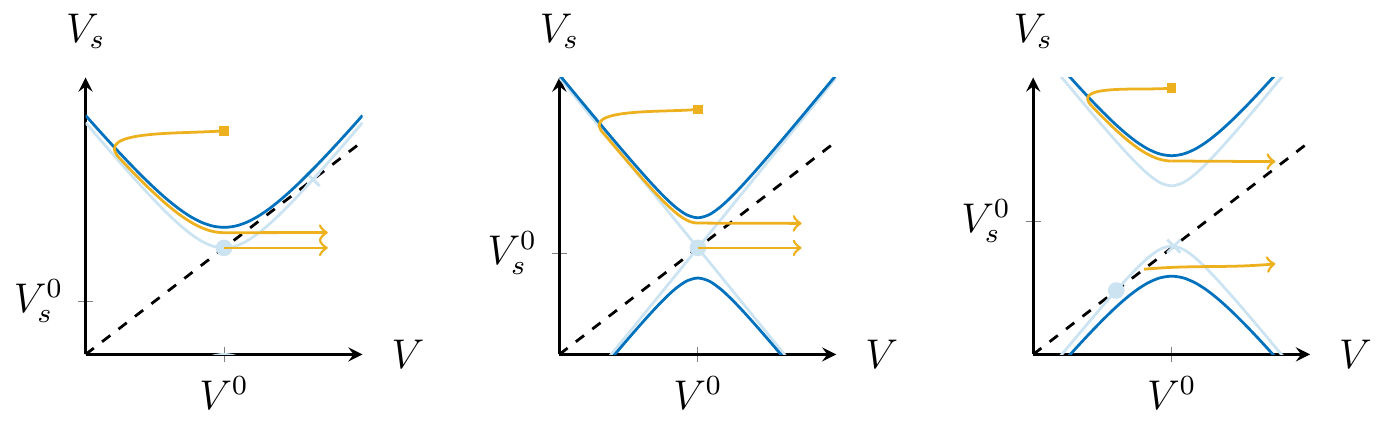}{.3}{.3}{Modulation of the excitability type in the MQIF model. Changing $V_s^0$ in the MQIF model results in different types of excitability: Type II for $V_s^0 < V^0$ (left), Type I for $V_s^0 = V^0$ (centre) and Type II*~\citep{Drion2015a} for $V_s^0 > V^0$ (right). The phase portraits show the $V$-nullclines just before (light blue) and after the bifurcation (dark blue). The $V_s$-nullclines are drawn as dashed lines. The stable fixed points are represented by filled circles, the saddle points by crosses. The possible trajectories are drawn in orange, with the reset points represented by squares.}

We can regard the MQIF model as a local approximation of the model in the present paper around a transcritical singularity. This singularity is identified in the model \eqref{eq:VC-analysis-NLIF-1}--\eqref{eq:VC-analysis-NLIF-3} by finding the point $(V^0,V_s^0)$ for which
\begin{align}
    \frac{\partial I_\text{ion}(V,V_s)}{\partial V} = \frac{\partial I_\text{ion}(V,V_s)}{\partial V_s} = 0\label{eq:VC-analysis-TC},
\end{align}
together with the condition that the determinant of the Hessian at $(V^0,V_s^0)$ should be negative (or zero, requiring further investigation). $I_\text{ion}(V^0,V_s^0)$ is then the current offset necessary to have the transcritical bifurcation occur at the same value of $I_\text{app}$. The values of $\bar{g}_f$ and $\bar{g}_s$ are simply the elements on the diagonal of the Hessian at $(V^0,V_s^0)$.

\subsection{Connection to dynamic input conductances}
\label{sec:VC-analysis-connection-DIC}

The analysis method of this paper has some similarities with that of the dynamic input conductances (DICs) introduced in \citet{Drion2015}. Both methods group the contributions of different ion channels into a fast, slow and ultraslow timescale. Nonetheless, a crucial difference between the methods is that the DICs are differential properties: they represent the change in current with an infinitesimal change in voltage. Instead, this paper considers the ion current itself as a function of voltages in different timescales. This is important because infinitesimal voltage clamp steps are hard to perform experimentally. In contrast, the voltage-clamp simulations considered in this paper could in principle be replaced by actual voltage-clamp experiments.

The other main difference between both methods is that DICs are defined around a steady state: all variables are at their steady-state value for the voltage $V$ at which the DIC is calculated. The function $I_{\text{ion}}(V,V_s,V_{us})$ in this paper, on the other hand, allows the voltages in each timescale to be different, thus capturing the transient behaviour of the total ionic current. This is an important difference, because it allows the method of this paper to identify an integrate-and-fire neuron model, which is not possible from DICs.

While not identical, because of the way variables are divided over timescales, a property similar to DICs can be derived from $I_{\text{ion}}(V,V_s,V_{us})$ as follows:
\begin{align}
  \label{eq:VC-analysis-DICs}
  g_f(V) &= \left.\frac{\partial I_{\text{ion}}(V_f,V_s,V_{us})}{\partial V_f}\right|_{V_f=V_s=V_{us}=V}\\
  g_s(V) &= \left.\frac{\partial I_{\text{ion}}(V_f,V_s,V_{us})}{\partial V_s}\right|_{V_f=V_s=V_{us}=V}\\
  g_{us}(V) &= \left.\frac{\partial I_{\text{ion}}(V_f,V_s,V_{us})}{\partial V_{us}}\right|_{V_f=V_s=V_{us}=V},
\end{align}
where $g_f$, $g_s$ and $g_{us}$ are the equivalent of the fast, slow and ultraslow DICs respectively. Since all voltages are taken as equal, it is clear that the DICs only reveal a subset of the information provided by $I_{\text{ion}}(V,V_s,V_{us})$.

\subsection{Connection to I-V curve analysis}
\label{sec:VC-analysis-connection-IV}

\citet{Ribar2019} propose a similar model of the total ionic current from I-V curves in different timescales. The main difference between the models is that the current functions in \citet{Ribar2019} are univariate. The current $I_{\text{ion}}$ is then a sum of functions of $V$, $V_s$ and $V_{us}$. The I-V curves are defined as the sum of all functions acting on their specific timescale and those faster, e.g. $I_f(V)+I_s(V)$ for the slow I-V curve.

\section{Conclusion}
\label{sec:VC-analysis-conclusion}

This paper introduced a method to obtain an integrate-and-fire model from a con\-duc\-tance-based model, by matching voltage-clamp and current-clamp responses.

The proposed multi-scale integrate-and-fire model has a simple structure, but retains a close connection to the physiology of the biophysical model.

The proposed method is applicable to arbitrary biophysical models under the key assumption that the kinetics of the gating variables can be grouped in a few distinct timescales.

\begin{appendices}

\section{Methods}
\label{sec:VC-analysis-methods}

\subsection{Software}
\label{sec:VC-analysis-software}

Simulations were performed in Python with the SciPy library using the equations stated in the text. Differential equations were solved using SciPy's backward differentiation formula (BDF) method. All figures were drawn using the Python packages Matplotlib and Tikzplotlib, and/or the \LaTeX packages PGF/TikZ and PGFPlots.

\subsection{Parameters of the biophysical models}
\label{sec:VC-analysis-CB-parameters}

The parameters of the biophysical models used in this paper take the original published values, except where a change of parameter is indicated. For the Connor-Stevens model \citep{Connor1977} these are the following: $C=\SI{1}{\micro\F\per\cm\squared}$, $\bar{g}_L=\SI{20}{\milli\siemens\per\cm\squared}$, $\bar{g}_{Na}=\SI{120}{\milli\siemens\per\cm\squared}$, $\bar{g}_K=\SI{20}{\milli\siemens\per\cm\squared}$,  $\bar{g}_A=\SI{47.7}{\milli\siemens\per\cm\squared}$, $V_L=\SI{-17}{\mV}$, $V_{Na}=\SI{55}{\mV}$, $V_K=\SI{-72}{\mV}$, $V_A=\SI{-75}{\mV}$.
The parameters for the model of \citet{Drion2018} are the means of the parameters used in their network simulations:
$C=\SI{1}{\micro\F\per\cm\squared}$,
$\bar{g}_L=\SI{0.055}{\milli\siemens\per\cm\squared}$,
$\bar{g}_{Na}=\SI{170}{\milli\siemens\per\cm\squared}$,
$\bar{g}_{K,D}=\SI{40}{\milli\siemens\per\cm\squared}$,
$\bar{g}_{Ca,T}=\SI{0.55}{\milli\siemens\per\cm\squared}$,
$\bar{g}_{K,Ca}=\SI{4}{\milli\siemens\per\cm\squared}$,
$\bar{g}_H=\SI{0.01}{\milli\siemens\per\cm\squared}$,
$V_L=\SI{-55}{\mV}$,
$V_{Na}=\SI{50}{\mV}$,
$V_K=\SI{-85}{\mV}$,
$V_{Ca}=\SI{120}{\mV}$,
$K_D=170$,
$k_1=0.1$,
$k_2=0.01$.

\subsection{Parameters of the integrate-and-fire models}
\label{sec:VC-analysis-IF-parameters}

All simulations and phase portraits of the integrate-and-fire models are based on the published biophysical model equations and parameters, unless stated otherwise in the text. The parameters of the integrate-and-fire models were found using the method described in Section \ref{sec:VC-analysis-init-opt}, with SciPy's Trust Region Reflective algorithm \citep{Branch1999} for the least-squares optimisation. The current clamp test data was generated using linearly decreasing currents for the Connor-Stevens model and a hyperpolarising step current for the model of \citet{Drion2018}. The data was sampled at \SI{100}{\kilo\Hz} and its transient was removed. The obtained values are given in Table \ref{tab:VC-analysis-IF-parameters} for each figure. The units of $C$, the time constants and the reset voltages are respectively \si{\micro\F\per\cm\squared}, \si{\ms}, and \si{\mV}.

\begin{table}[htbp!]
\centering
\caption[Parameters of the integrate-and-fire models]{Parameters of the integrate-and-fire models.}
\begin{tabular}{l c c c c c c c}
  \toprule
  Figure & {\bf $C$} & {\bf $\tau_f$} & {\bf $\tau_s$} & {\bf $\tau_{us}$} & {\bf $V_r$} & {\bf $V_{s,r}$} & {\bf $\Delta V_{us}$} \\
  \midrule
  \ref{fig:CS-VC-PP-2} \& \ref{fig:CS-VC-PP-gA} (centre) & $0.58$ & $0.022$ & $6.7$ & & $-40$ & $-25$ & \\
  \ref{fig:CS-VC-PP-gA} (left) & $0.86$ & $0.03$ & $3$ & & $-45$ & $-24$ & \\
  \ref{fig:CS-VC-PP-gA} (right) & $0.3$ & $0.027$ & $23$ & & $-35$ & $-24$ & \\
  \ref{fig:CS-VC-3ts} & $1.2$ & $0.037$ & $1.7$ & $2.8$ & $-40$ & $-20$ & $0$\\

  \ref{fig:Drion-VC-step}-\ref{fig:Drion-hybrid} & $0.82$ & $0.89$ & $4.3$ & $278$ & $-45$ & $7.5$ & $1.7$\\
  \bottomrule
\end{tabular}
\label{tab:VC-analysis-IF-parameters}
\end{table}

\end{appendices}

\section*{Acknowledgements}
The authors would like to thank Ilario Cirillo, Thiago Burghi, Luka Ribar and Christian Grussler for the helpful discussions.
TVP received a fees scholarship from the Engineering and Physical Sciences Research Council (https://www.epsrc.ac.uk) under grant number 1611337. Both TVP and RS were supported by the European Research Council (https://erc.europa.eu) under the Advanced ERC Grant Agreement number 670645. The funders had no role in study design, data collection and analysis, decision to publish, or preparation of the manuscript.

\bibliography{VC-analysis}

\end{document}